# Reversible Electrochemical Phase Change in Monolayer to Bulk MoTe$_2$ by Ionic Liquid Gating

**Dante Zakhidov[1], Daniel A. Rehn[2,3], Evan J. Reed[1], Alberto Salleo[1,*]**

[1]Department of Materials Science and Engineering, Stanford University, Stanford, CA 94305, USA
[2]Department of Mechanical Engineering, Stanford University, Stanford, CA 94305, USA
[3]Los Alamos National Laboratory, Los Alamos, NM 87545, USA

*Email: asalleo@stanford.edu

**ABSTRACT**

Transition metal dichalcogenides (TMDs) exist in various crystal structures with semiconducting, semi-metallic, and metallic properties. The dynamic control of these phases is of immediate interest for next generation electronics such as phase change memories. Of the binary Mo and W-based TMDs, MoTe$_2$ is attractive for electronic applications because it has the lowest energy difference (40 meV) between the semiconducting (2H) and semi-metallic (1T') phases, allowing for MoTe$_2$ phase change by electrostatic doping. Here we report phase change between the 2H and 1T' polymorphs of MoTe$_2$ in thicknesses ranging from the monolayer case to effective bulk (73nm) using an ionic liquid electrolyte at room temperature and in air. We find consistent evidence of a partially reversible 2H-1T' transition using in-situ Raman spectroscopy where the phase change occurs in the top-most layers of the MoTe$_2$ flake. We find a thickness-dependent transition voltage where higher voltages are necessary to drive the phase change for thicker flakes. We also show evidence of electrochemical activity during the gating process by observation of Te metal deposition. This finding suggests the formation of Te vacancies which have been reported to lower the energy difference between the 2H and 1T' phase, potentially aiding the phase change process. Our discovery that the phase change can be achieved on the surface layer of bulk materials reveals that this electrochemical mechanism does not require isolation of a single layer and the effect may be more broadly applicable than previously thought.

**Introduction**

Since the discovery of graphene[1], the field of 2D materials has rapidly grown with the introduction of new classes of 2D materials such as buckled, single-element Xenes[2] (silicene, germanene, tellurene), transition metal carbides and nitrides called MXenes[3,4] ($Ti_3C_2$, $Ti_4N_3$), and the transition metal dichalcogenides[5,6] ($MoS_2$, $WTe_2$, $TaSe_2$). The study of these materials has been motivated by their attractive thermal, electronic, optical, and catalytic properties that can be tuned not only by size, composition, and functionalization, but also more fundamentally altered by their crystal structure or phase. 2D materials often exhibit polymorphism, with crystal phases in a given material having vastly different electronic structures that allow for phenomena such as ferroelectricity, piezoelectricity, charge density waves, superconductivity and their corresponding applications.[1–6] Particularly, the ability to switch between semiconducting and metallic phases in the transition metal dichalcogenides (TMDs) is of immediate interest for next-generation electronics such as phase change electronic memories and other applications due to their unique properties, smaller device footprint, and potentially lower energetic switching costs[7,8].

The majority of research on phase equilibria has focused on the group VI TMDs ($MoX_2$, $WX_2$) between the main polymorphs which are the trigonal-prismatic coordinated, semiconducting 2H phase; the octahedral coordinated, metallic 1T phase; and the distorted octahedral-coordinated, semi-metallic 1T' phase (Figure 1a)[9]. Phase control and phase change have been reported in group VI TMDs by means of alkali-metal intercalation[10], alloying[11,12], AFM-tip induced strain[13], joule heating[14], thermal treatment[15,16], and electrostatic gating[17]. Out of those binary TMDs, $MoTe_2$ has the lowest energy difference between the semiconducting 2H and the semi-metallic 1T' phase with the potential to further reduce the energy difference by alloying with W[9,18]. Furthermore, computational work on monolayer $MoTe_2$ showed that the low energy difference between 2H and 1T' makes the phase transition obtainable by charging via electrostatic gating[19]. Electrostatic phase change by means of field-induced charge doping remains the most attractive switching mechanism for device technologies because of its potential to be reversible, energy-efficient, and fast. The density functional theory (DFT) studies were first experimentally confirmed by Wang et al. who showed reversible phase change in monolayer $MoTe_2$ using a gated ionic liquid at 220K under vacuum[17].

We extend the study by Wang *et al.* to demonstrate phase change in all thicknesses of $MoTe_2$, ranging from the monolayer case to bulk-like (73nm), using ionic-liquid gating with N,N-diethyl-N-(2-methoxyethyl)-N-methylammonium bis(trifluoromethylsulphonyl-imide (DEME-TFSI) at room temperature and in air. We find consistent evidence of a partially reversible 2H-1T' transition using in-situ Raman spectroscopy and demonstrate the appearance of two additional 1T' Raman modes that are widely reported in pristine 1T' but previously not reported in 1T' $MoTe_2$ produced by phase change from the 2H phase. We empirically find that the transition voltage increases with thickness, suggesting that higher charge densities are necessary for thicker flakes. Ultimately, we demonstrate that our phase change is not purely electrostatic but also electrochemically mediated, likely by the creation of Te vacancies during the gating process. Te vacancies have been reported to lower the energy difference between the 2H and 1T' phases under some conditions[15,20–22], and likely decrease the doped charge density necessary to switch MoTe$_2$ flakes thicker than a monolayer. Additionally, polarization-resolved Raman measurements suggest that the 1T' phase produced by our experimental conditions is polycrystalline, with domains forming heterogeneously along different 2H crystallographic directions. These results display MoTe$_2$ phase change in a variety of new and practical conditions, providing promise for MoTe$_2$ in phase change memories. They also suggest that controlled vacancy creation or flake thickness can be a valuable knob in tuning the transition voltage for phase change memories.

**Results and Discussion**

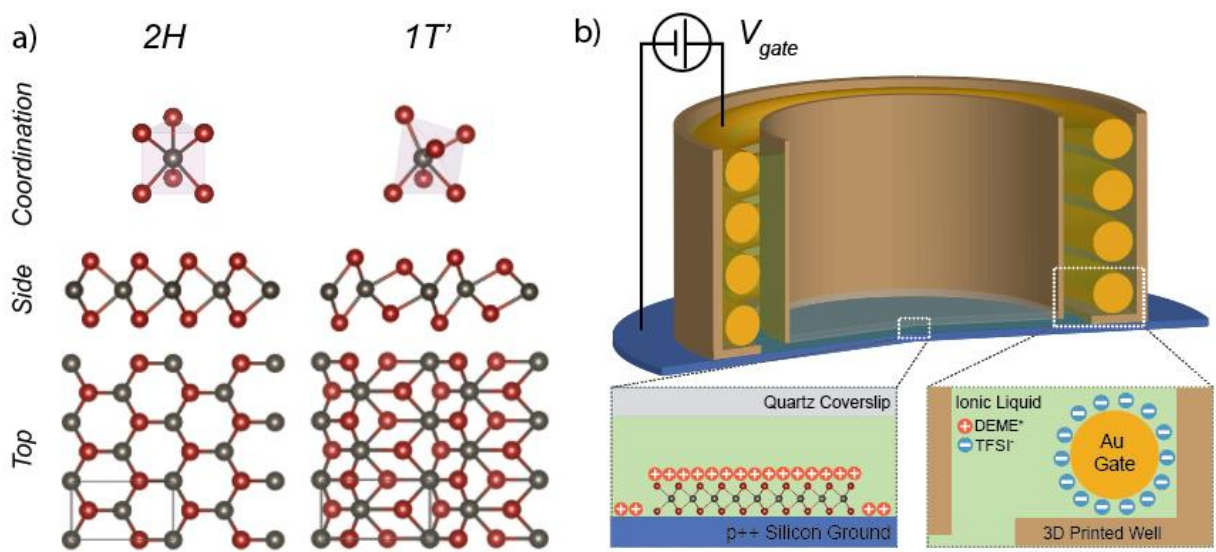

**Figure 1 Crystalline Phases of MoTe$_2$ and Ionic Liquid Gating Structure**. **a)** Crystal structure of 2H and 1T' MoTe$_2$ phases. The grey atoms represent molybdenum and the red atoms represent tellurium. The 2H phase is semiconducting with a trigonal prismatic coordination and the 1T' phase is semi-metallic with a distorted octahedral coordination. **b)** Measurement cell and configuration used to manipulate the MoTe$_2$ phase. The MoTe$_2$ is exfoliated onto a p++ Silicon substrate which doubles as an electrode for charge carrier injection. A plastic 3D printed well houses the gate and the DEME-TFSI ionic liquid. The surface area of the gate is designed in such a way as to not limit the gating process. A thin quartz coverslip attached to an inner well is used to minimize the volume of ionic liquid along the Raman laser path.

We investigate the phase change between the semiconducting 2H and semi metallic 1T' phase of MoTe$_2$ for flake thicknesses ranging from the monolayer to ~104 layers (73nm). Charge injection into MoTe$_2$ is realized using a gated DEME-TFSI ionic liquid because it has been shown to be able to generate a sufficiently large electron charge density (>1 x 10$^{14}$ cm$^{-2}$)[17,19] to induce the phase change (Figure 1b, see Methods). Since the 2H and 1T' phase exhibit different crystal structures, they also have distinct phonon modes. As a result, we probe the phase change using in-situ Raman spectroscopy. Notation of the 2H and 1T' MoTe$_2$ Raman modes are clarified in the supplementary material (SI Table 1).

Bulk Phase Change

Figure 2a shows the Raman spectra during a full gating cycle of a bulk-like, 60nm MoTe$_2$ flake. At 0V, the characteristic out of plane vibrational mode, 2H A$_{1g}$, at 173 cm$^{-1}$ and the in-plane vibrational mode, 2H E$_{2g}$, at 234 cm$^{-1}$ are observed. The low intensity of the A$_{1g}$ peak is characteristic of bulk flakes where the out-of-plane vibrations are dampened by many-layer interactions[23]. The broad peak at 120 cm$^{-1}$ and the range of peaks from 270-350 cm$^{-1}$ are due to the DEME cation and TFSI anion of the ionic liquid, respectively[24]. As the voltage is increased up to 4.0 V, no significant peak shifts or peak intensity changes are observed in the 2H A$_{1g}$ and 2H E$_{2g}$ peaks.

At 4.0V the critical voltage necessary to induce phase change is attained, which is marked by the appearance of an initially small 1T' A$_{1g}$ peak at 168 cm$^{-1}$ as highlighted in Figure 2b. The position of this 1T' Raman peak is consistent with the previous report of electrostatic phase change and corroborates those findings[17]. This voltage marks the beginning of nucleation of 1T' domains in the 2H flake. A large increase in the 1T' A$_{1g}$ intensity is observed at 4.2V, accompanied by a decrease in intensity and broadening of the 2H E$_{2g}$ peak. Concurrently, we observe the rise of additional peaks at ~78 cm$^{-1}$ and ~90cm$^{-1}$, which we attribute to the A$_{1g}$ and B$_{1g}$ modes of 1T'[25]. These peaks have not been previously observed in 2H to 1T' MoTe$_2$ phase change experiments but are consistently reported as prominent 1T' peaks in the DFT and CVD-growth literature (SI Table 2).

Fundamentally, the linewidths of the Raman modes are inversely related to the phonon lifetime, with broader linewidths indicating faster phonon decay, generally attributed to increased number of defects.[26] The broad nature of the 1T' 78 and 90 cm$^{-1}$ peaks as well as the 2H E$_{2g}$ peaks suggest that upon ionic liquid gating the 1T' phase as well as the remaining 2H phase become defective. A discussion of the Raman peak center and linewidths as well as their values are included in Supplementary Figures 1 and 2.

Upon removing charge, several effects are observed. At 3V, the 78 and 90 cm$^{-1}$ peaks disappear, corroborating the fact that they arise from the gate-induced phase transformation and do not come from parasitic chemical reactions. The 168 1T' peak on the other hand persists but lowers in intensity. This decrease in the 1T' peak intensity coincides with a strong increase in the intensity of the 2H E2g peak and A1g peaks, beyond their original intensity. Such a strong intensity increase is not completely understood at this stage but is possibly due to a SERS effect (see Supplementary Figures 3 and 4) and was not reported in monolayer 2H to 1T' phase change experiments[17]. The strong increase in intensity makes the 2H A$_{1g}$ peak at 172 cm$^{-1}$ more visible as compared to the initial state and this signal most likely carries a strong contribution from 2H layers underneath the 1T'. At 1V, 1T' MoTe$_2$ transforms back to the 2H phase with the addition of a shoulder off the ~172 cm$^{-1}$ 2H $A_{1g}$ peak. We attribute this shoulder to a potential distribution of 2H domains with varying strains and defect densities that arise from phase change cycling. In summary, the phase change process appears to be partially reversible in thick flakes.

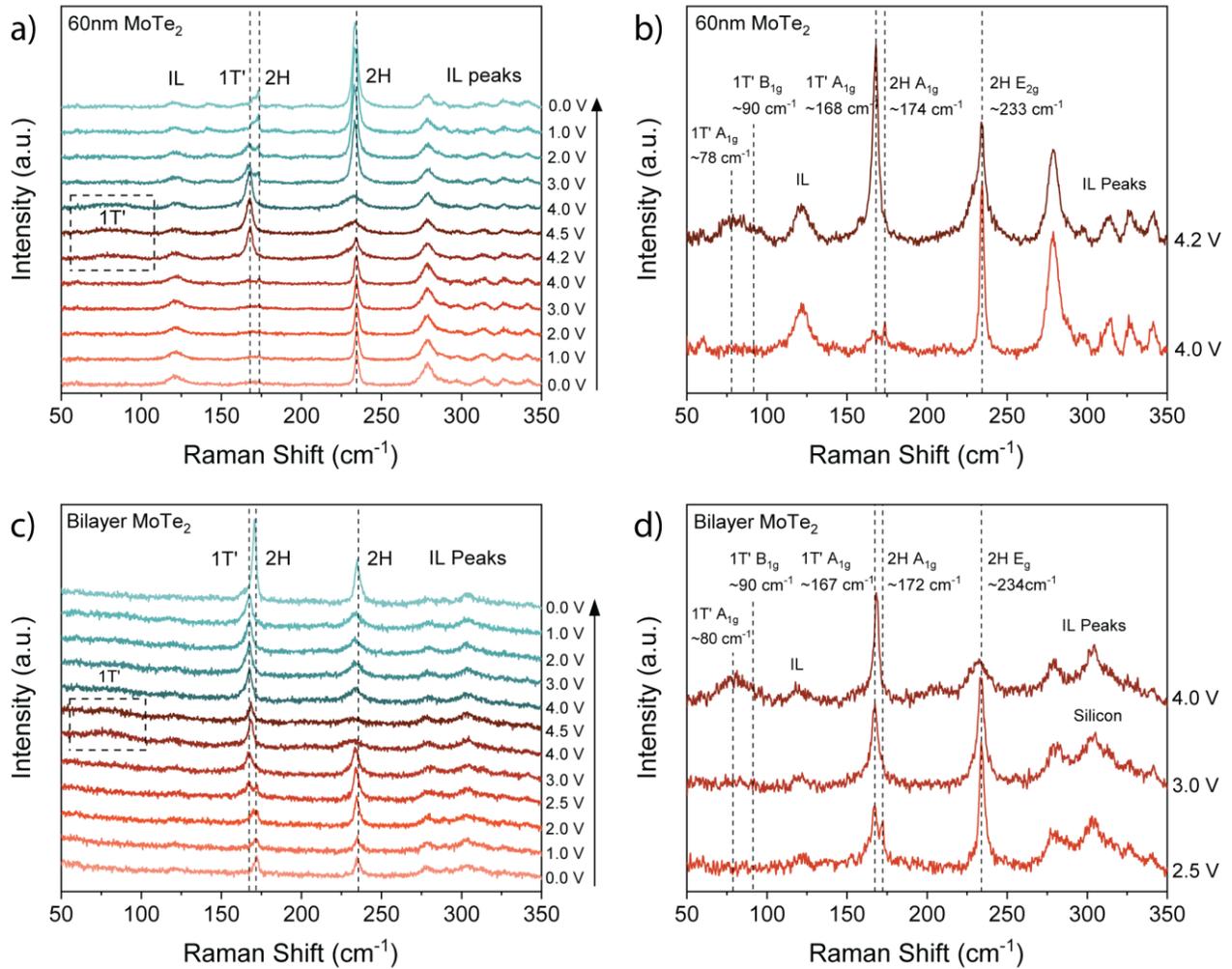

**Figure 2. Raman Characterization of the 2H to 1T' MoTe₂ Phase Transition**. **a,b)** In-situ Raman progression of a bulk-like 60nm MoTe₂ flake under ionic liquid gating from 0V to 4.5V to 0V. Figure 2b highlights the transition voltage at 4.0V in Figure 2a and shows the appearance of the 1T' peaks at ~78 cm$^{-1}$, ~90 cm$^{-1}$, and ~168 cm$^{-1}$. **c,d)** In-situ Raman progression of a bilayer MoTe₂ flake under ionic liquid gating from 0V to 4.5V to 0V. The spectra are qualitatively similar to those observed in the bulk case. Figure 2d also shows the 2.5V transition voltage in Figure 2c and the coexistence of the 1T' and 2H peaks at this voltage.

Bilayer Phase Change

Building up towards the bulk regime described above, a similar pattern of phase change is observed in bilayer MoTe₂ as seen in Figure 2c and 2d. The 2H $A_{1g}$ and 2H $E_{2g}$ peaks start with nearly identical intensities, which is characteristic of bilayer MoTe₂ [27]. In contrast to the thicker 60nm MoTe2 flake, phase change in the bilayer MoTe₂ occurs at 2.5V where co-existence of the

1T' $A_{1g}$ and 2H $A_{1g}$ is observed (Fig. 2d). Unlike the bulk case, the 2H $E_{2g}$ peak does not drop in intensity at this transition voltage. It isn't until 4V that the intensity of the 2H $E_{2g}$ drops significantly, the linewidth broadens, and a broad convolution of the 1T' peaks at 80 cm$^{-1}$ and 90 cm$^{-1}$ appear. Upon discharging the flake, the flake reverts to the 2H phase only at 0V, indicating that a strong kinetic barrier induces hysteresis, as previously reported[17]. At 0V, the intensities of the 2H peaks are larger than prior to gating, which we attribute again to the potential SERS effect. Alternatively, since the intensity of the 2H $A_{1g}$ peak is higher than that of the 2H $E_{2g}$ peak, which is characteristic of monolayer MoTe$_2$, this result could be interpreted as thinning of the bilayer to monolayer. In either case, a similar pattern of phase change is observed in both bulk-like and bilayer MoTe$_2$

Discussion of Phase Change Mechanism

Throughout the bilayer and bulk experiments discussed above, the 2H $E_{2g}$ signal never entirely disappears, indicating coexistence of the 2H and 1T' phases. The coexistence suggests two possible scenarios. First, the phase change could only occur on the top or top-few layers and the 2H Raman signature originates from underlying layers. Alternatively, the phase change is not complete even on the topmost layers and the Raman measurement probes both 2H and 1T' domains on the surface. We hypothesize that what is observed is a combination of both effects.

To investigate whether charge can be transferred from the ionic liquid to layers below the top layer, we perform density functional theory calculations of a system consisting of a lithium atom sitting on top of a bilayer of 2H-MoTe2 (Figure 3). The lithium atom is used as a proxy for the ionic liquid in order to investigate doping of the bilayer with electrons. The quantity of interest is the amount of charge that moves from the lithium atom to the bottom layer. To investigate this, we compute the fractional charge $N(z)$ of the excess electron coming from the lithium atom at location $z$ in the cell for three different lithium atom positions above the bilayer, plotted on the x axis of Figure 3,

$$N(z) = \int_0^z \left( n_{\text{Li}+\text{MoTe}_2}(x', y', z') - n_{\text{MoTe}_2}(x', y', z') \right) dx'dy'dz'$$

Here, n$_{\text{Li}+\text{MoTe2}}$ is the charge density of the Li + bilayer system and n$_{\text{MoTe2}}$ is the charge density of the bilayer. Regardless of the position of the lithium atom with respect to the surface, we find that no appreciable charge density moves past the first layer. Notably, the charge density associated with the excess electron coming from the lithium atom is around 0.25 e/f.u. MoTe$_2$ in these calculations, much higher than the charge density predicted[19] and measured[17] as necessary to cause the transformation, around 0.05 e/f.u. MoTe$_2$. While an idealized case, the DFT results dispel the possibility of a 2H to 1T' transition in all layers of the sample and supports the idea of a top or top-few layer transition. In the case of a few-layer transition, the DFT studies show charge donation from the top layer to layers below are unlikely to occur in the pristine bilayer case. This suggests that charge donation to layers below must be mediated by local defects in the sample, such as intercalants between layers that mediate charge transfer. Another possibility is that the Te vacancies generated during the gating process allow for mobile Te atoms to mediate charge transfer between layers. In either case, it is unlikely for charge to penetrate more than a few layers below the surface.

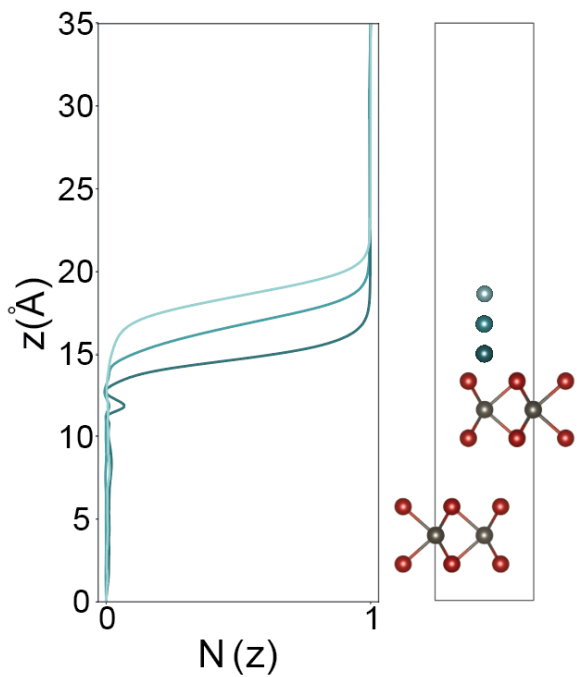

**Figure 3. Location of Electrostatically Doped Charge Density.** Varying the position of a lithium atom on a MoTe$_2$ layer, we find that the DFT-computed excess charge stays almost entirely in the first layer.

Experimentally, we hypothesize that the newly reported ~78 cm$^{-1}$ 1T' A$_{1g}$ and ~90 cm$^{-1}$ B$_{1g}$ modes might be due to a few-layer phase change, where additional 1T' layers increase the Raman signal and allow the peaks to be detected, compared to the monolayers. This hypothesis is based on the observation that new A$_{1g}$ and B$_{1g}$ 1T' modes do not always appear and disappear concurrently with the main ~168 cm$^{-1}$ A$_{1g}$ 1T' peak, which indicates a multi-step process. For example, in the bilayer case, the phase change occurred at 2.5V but the appearance of the new 1T' modes and broadening of the 2H E$_{2g}$ did not occur until 4V. Additionally, assuming the phase change occurred in both layers, we still observe that the 2H E$_{2g}$ peak does not fully disappear. This result implies that the phase change is not complete, with remaining regions of 2H MoTe$_2$ in the top or bottom layer.

We also consider the possibility that the phase change is driven by intercalation of the DEME cations into the MoTe$_2$ flakes. However, using ex-situ XRD, we don't find any evidence of intercalation (Supplementary Figure 5). Thus, we suggest that the phase change happens in the topmost layers due to the formation of an electric double layer and the extent of the phase change in the lateral direction is not complete.

## Thickness Dependence

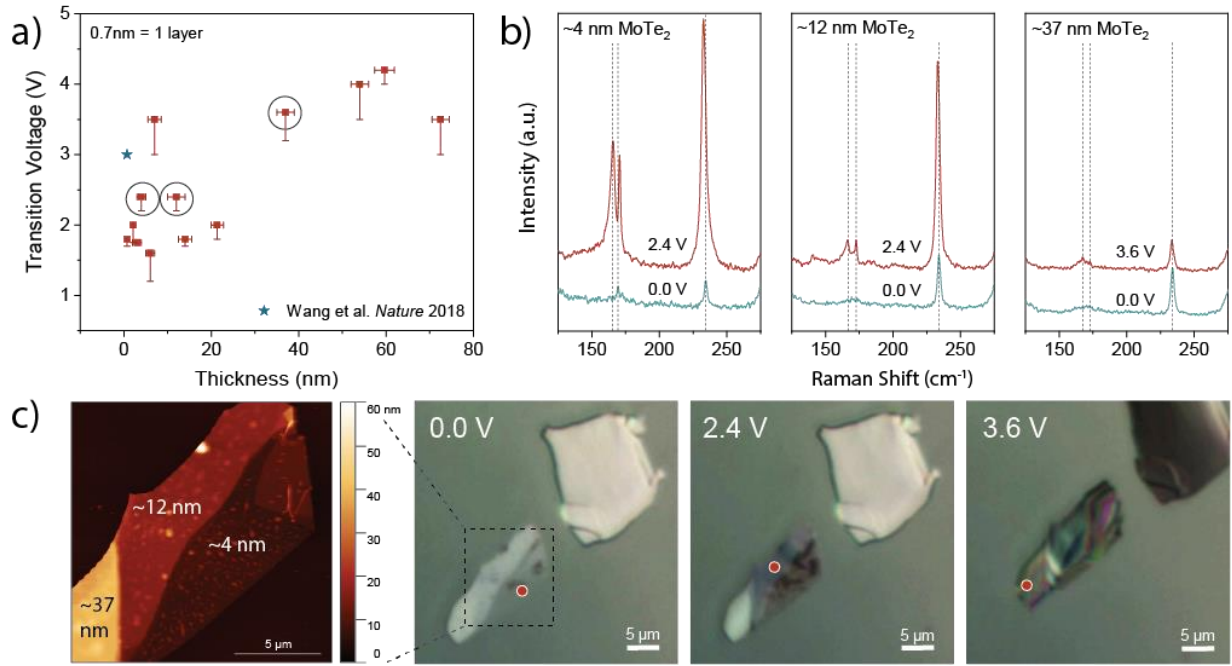

**Figure 4. Thickness Dependence on the Transition Voltage. a)** Comparing the 2H to 1T' transition voltage in MoTe$_2$ flakes of different thicknesses taken from in-situ Raman measurements. Wang et al.'s work on monolayer MoTe$_2$ is shown as reference. The y-axis error bars represent voltage steps used in the gating experiments and the x-axis error bars represent uncertainties from AFM measurements. The grey circles highlight the experiments show in figure 4b and 4c **b)** The initial and transition voltage Raman spectrum of MoTe$_2$ flakes of different thicknesses, all plotted on the same intensity scale. These spectra correspond to the regions in the optical images in figure 4c and are marked by a green dot **c)** AFM image showing the ~4, ~12, and ~37 nm regions that are depicted in the three optical images. The optical images show a color change on the regions of the flake associated with their transition voltage.

In addition to the bilayer and bulk-like case, phase change was observed in MoTe$_2$ flakes of varying thicknesses (Figure 4a) where we observe that thicker flakes require a larger voltage to induce phase change. The transition voltage is defined as the voltage where the 1T' peak is first detected in the in-situ Raman measurements. We highlight three of those thicknesses and show the Raman spectra at the transition voltage in Figure 4b and the AFM and optical images of the corresponding flake in Figure 4c. In the Raman spectrum, we see that the 4nm and 12nm regions of the flake both undergo phase change at 2.4V with the appearance of the 1T' peak. Yet again, we see a sharp increase in the intensity of all the peaks, which we attribute to a surface enhancement effect that is not completely understood. In the 37nm region of the flake, phase change only occurs at 3.6V and we do not see the surface enhancement effect, making the phase change difficult to distinguish. The optical images, taken during the measurement, serve to illustrate the phase change with distinct coloration occurring at the transition voltages. The images visually confirm the thickness dependence on the transition voltage, with the 4nm and 12nm regions changing color at 2.4V but the 37nm region only changing color at 3.6V. A video of the phase change process on the flake shown in Figure 4c is shown in Supplementary Video 1.

In the case of few-layer flakes, where interlayer interactions can lead to Davydov peak splitting, it is important to distinguish the gate-induced 1T' peak from a potential Davydov peak. In Supplementary Figure 6, we show how the spectral features of the 1T' peak are distinct from Davydov peaks, disproving potential concerns of peak misidentification. Given the observed phase change phenomena in a variety of thicknesses, we hypothesize that the phase change is electrochemically mediated through the creation of Te vacancies as discussed below.

Electrochemical Phase Change by Creation of Te Vacancies

The coloration of the $MoTe_2$ flakes (Fig. 4c) during the gating process indicates the possibility of electrochemical reactions on the surface of the $MoTe_2$. Indeed, Raman characterization of the $MoTe_2$ flakes upon gating revealed the presence of Te metal either on or around the $MoTe_2$ flakes at the end of the gating process. Figure 5 shows a series of Raman maps highlighting the presence of a Te metal film next to a $MoTe_2$ flake. Representative spectra from the Raman maps are plotted in Figure 5e displaying the characteristic Te metal peaks[28,29] at ~120 cm$^{-1}$ and ~140 cm$^{-1}$ as well as the 2H $E_{2g}$ $MoTe_2$ and Silicon peaks. Because the p++ Si electrode is used as a universal contact, electrically connecting every flake on the substrate, we hypothesize that the Te metal film plates out of the ionic liquid and that the Te atoms potentially come from all $MoTe_2$ flakes on the substrate and not solely from the flake in the image. As a result, the Te metal is not favored to plate out in any specific location and can form far away from the location of measurement. Consequently, we do not always observe the presence of Te metal and are unable to track its formation to make any correlations to the time or rate of Te vacancy creation. Thus, our measurements do not indicate whether vacancies are always present when the phase change is observed. Furthermore, we cannot state if there is an intermediate voltage regime where phase chase is observed without vacancy formation. Nevertheless, this measurement strongly suggests the dissociation of the Mo-Te bond by electrochemical reaction with the ionic liquid, possibly resulting in a large concentration of Te vacancies left in the flakes.

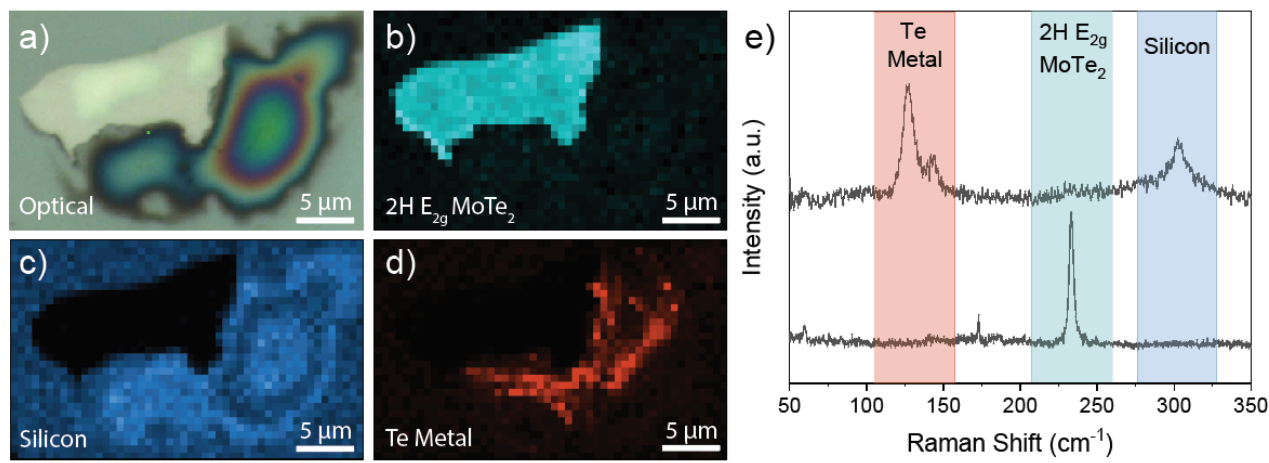

**Figure 5. Precipitation of Tellurium Metal. a)** Optical image of a $MoTe_2$ flake after most of the ionic liquid was removed at the end of a gating cycle. Raman maps of **b)** the $MoTe_2$ flake as shown through the 2H $E_{2g}$ peak, **c)** the Silicon substrate, and **d)** the Te metal as shown through the characteristic ~120 and ~140 cm$^{-1}$ peaks. **e)** A legend to show what Raman peaks are plotted in the Raman maps.

Importantly, Te vacancies can play a crucial role in reducing the energy difference between the 2H and 1T' phases due to their role as n-type dopants.[21] Reported DFT calculations suggest that $MoTe_2$ flakes with Te vacancy concentrations greater than 2% are more stable in the 1T' phase than in the 2H phase.[21] Experimentally, studies on the CVD growth of $MoTe_2$ found that tellurization of a $Mo/MoO_3$ film resulted in the 1T' phase in Te-deficient conditions while it favored the 2H phase in excess Te conditions.[15,20] Furthermore, Keum et al. demonstrated that the transformation temperature from the 2H to the 1T' phase decreased to ~650 °C from ~820 °C in Te-deficient $MoTe_2$.[22] Computational and experimental evidence support the conclusion that Te-deficient $MoTe_2$ favors the 1T' phase.

To further investigate the role of Te vacancy formation, we explore the effect of Te vacancies on the Raman modes used to identify 1T' $MoTe_2$ through DFT-based methods. We compute the phonon modes of pristine and ~3% Te-deficient $MoTe_2$ (see Methods section). The results, indicate that even with a relatively high 3% Te vacancy concentration, the calculated Raman active $A_g$ mode of the 1T' phase at 163 $cm^{-1}$ remains remarkably similar in frequency and vibrational pattern to the $A_g$ mode of pristine $MoTe_2$ (Supplementary Figure 7). This suggests that the Raman spectra found in experiments would not be sensitive enough to detect Te vacancy formation via a frequency shift in the active $A_g$ mode.

These observations lead us to hypothesize that the mechanism of phase change in $MoTe_2$ flakes in our experiments has both an electrochemical and an electrostatic component. In fact, it is now understood that transistor control through electrolyte gating is not fully explained by electrostatic induction of carriers and can proceed via electrochemical mechanisms.[30] For example, electrochemical mechanisms evidenced by the formation of oxygen vacancies have been reported in a number of oxide systems such as $VO_2$, $La_{1-x}Sr_xCoO_{3-\delta}$, and $SrTiO_3$.[30] In our case, we have a chalcogenide system where we speculate that the electrochemical creation of Te vacancies during electrolyte gating lowers the energy difference between 2H and 1T' and the simultaneous electrostatic doping drives the transition.

Reversibility

Even though the phase change is mediated by tellurium vacancies, the phase transformation process retains an electrostatic component. The reversible nature of the phase change in the 12nm MoTe$_2$ flake previously discussed is displayed in Figure 6a. The 1T' Ag peak disappears upon decreasing the voltage to 0V and re-appears upon gating back to 2.4V. Since this is a thicker, 12nm flake, the 2H $A_{1g}$ peak is known to be hard to distinguish but the wider range scan shows the presence of the 2H $E_{2g}$ to confirm the 2H nature of the flake. Figure 6b shows reversible phase switching of a monolayer MoTe$_2$ flake. The characteristic signature of monolayer 2H MoTe$_2$ where the $A_1'$ is higher in intensity than the E' peak is visible.[23] The sharp 2H $A_1'$ peak disappears and the broad 1T' $A_g$ peak appears upon gating to 1.8V. The 2H E' peak also disappears, indistinguishable from the noise. After cycling back to 0V, we see the 2H peaks return, but differing in intensity and peak position. Additionally, a shoulder on the 2H $A_1'$ peak appears (Figure 1b inset). These observations point to possible defect formation and the only partial reversibility of the process.

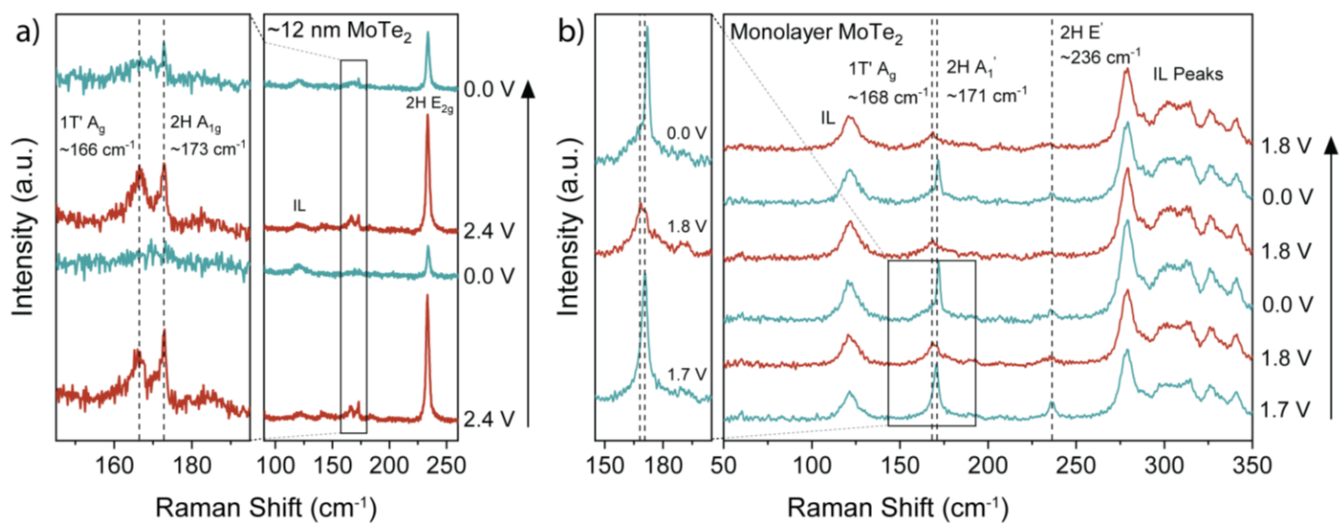

**Figure 6. MoTe$_2$ Phase Cycling. a)** Reversibility in 12nm MoTe$_2$ where the 2H to 1T' transition happens at 2.4V. The inset highlights the 2H and 1T' peaks while the zoomed-out portion shows there is an intensity enhancement at 2.4V. **b)** Reversibility in monolayer MoTe$_2$ with the transition voltage at 1.8V. The peaks are named according to crystal symmetry convention as clarified in the supplementary information. The inset shows one cycle of phase switching and highlights the shoulder of the 2H $A_1'$ peak after reducing the voltage to 0.0 V.

Heterogeneous Nucleation of 1T' Domains

The breadth of the 1T' Raman peak and the broadening of the 2H peaks upon cycling suggest that the single crystal nature of the material is perturbed during the gating process. In order to test the in-plane crystalline texture of the gate-induced 1T' MoTe$_2$, we investigated the polarization dependence of the 1T' $A_{1g}$ peak. Symmetry analysis of the 1T' $A_{1g}$ mode shows that the 1T' A$_{1g}$ mode has anisotropic scattering components that result in a two-lobe pattern in polarized Raman measurements (Fig. 7a). In contrast, in the gate-induced 1T' there is no polarization dependence, suggesting that the resulting 1T' is polycrystalline with domains forming along different crystallographic directions (Fig 7b). This result contrasts with work by Wang et al, who demonstrated the characteristic two-lobe pattern in their gate-induced 1T' at 220K and in vacuum[17]. We speculate that the loss of polarization dependence is due to the formation of Te vacancies and the electrochemical contribution to the transformation process observed here, which allow the 1T' to nucleate isotropically. This hypothesis is supported by a recent UHV-STM report of 2H MoTe$_2$ after a 400 °C thermal treatment which showed the presence of 60° inversion domain boundaries and was explained by Te vacancy creation[31]. We tested this assumption by performing the same gating conditions on the pristine, exfoliated 1T' flake. The initial polarization dependence of the exfoliated 1T' is lost after gating past 3V. These results indicate that the electrolyte gating process allows the MoTe$_2$ surface to reorganize which explains why the gate-induced 1T' does not exhibit any polarization dependence.

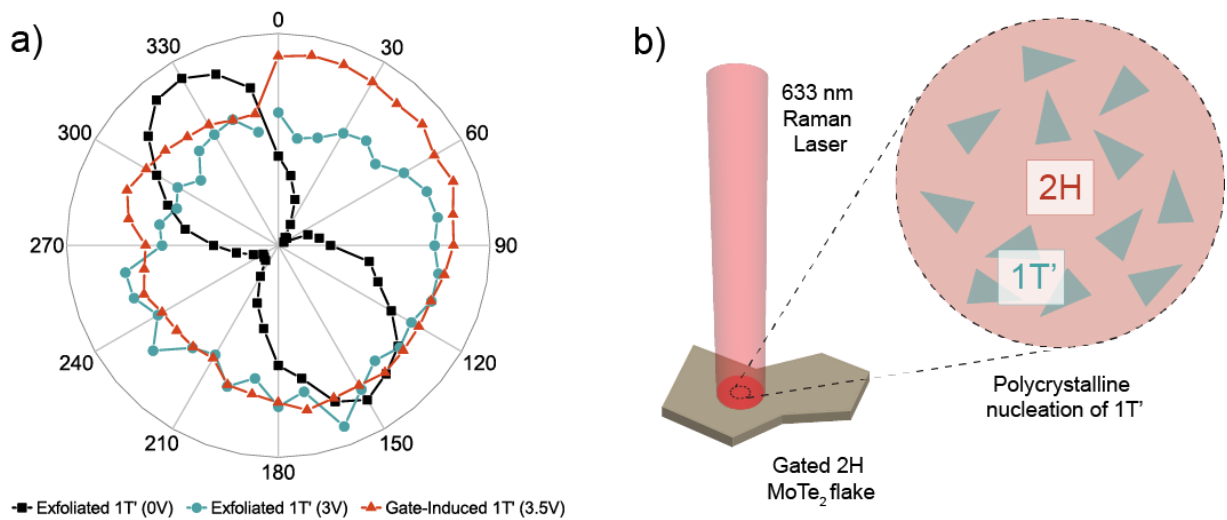

**Figure 7. Isotropic 1T' Nucleation.** a) Comparing polarization-resolved Raman of the 1T' A$_{1g}$ peak between pristine, exfoliated 1T' at 0V and 3V and gate-induced 1T' from a 2H flake. The exfoliated 1T' was gated in the same gating conditions as all other experiments. b) Schematic depicting the polycrystalline nucleation of 1T' domains in a 2H layer that would lead to the loss of polarization dependence seen in the gate-induced 1T'(3.5V) of part a.

**Conclusion**

We report evidence of 2H to 1T' phase change in MoTe$_2$ flakes upon ionic liquid gating at room temperature and ambient conditions in flake thicknesses ranging from monolayer to bulk-like

(73 nm). We find that the phase change is partially reversible through electrostatic doping and also electrochemically-mediated by the creation of Te vacancies which is supported by the observation of Te metal formation. The process leaves a mixture of phases as not all of the 2H in a given layer is transformed to 1T', with the transformation possibly occurring beyond the first layer. Furthermore, the 1T' domains are nucleated with varying crystalline orientations which results in a loss of the polarization dependence of the 1T' Raman spectrum.

This is the second report of electrically controlled phase change in $MoTe_2$. While the previous report showed the phase change at 220K and in vacuum, our observation of reversible phase change at room temperature and ambient conditions is relevant for implementation of $MoTe_2$ in phase change memory devices. This work also highlights the simultaneous activation of an electrochemical and an electrostatic switching mechanism as well as the important role of Te defects and the necessity of better understanding them. Pre-treatment of $MoTe_2$ devices to induce Te defects by annealing, plasma bombardment, or electrochemical dissociation may be a way to tune the transition voltage into a desirable range for phase change memory applications.

**Methods**

Exfoliation of 2H and 1T' $MoTe_2$ Flakes

2H and 1T' $MoTe_2$ flakes were mechanically exfoliated from commercially-purchased bulk crystals from HQ Graphene using 3M 810 Magic Tape or Blue Medium 18074 Thermal Release Tape from Semiconductor Equipment Corp. All exfoliations were done in air and the peel back of the tape was done quickly. A quick rip was qualitatively found to produce higher yields of few-layer $MoTe_2$ flakes. The exfoliated flakes were either directly applied to a p++ silicon substrate for testing or first transferred to a PDMS stamp to be then stamped onto the silicon substrate. When transferring from PDMS to the silicon substrate, the PDMS was put into contact with the substrate and then heated to 40 °C for 2 minutes before being ripped off.

Thickness Determination of $MoTe_2$ Flakes

The thickness of the $MoTe_2$ flakes was determined by a combination of optical microscopy, AFM, and Raman spectroscopy. First, promising monolayer to few-layer regions were identified by microscope based on optical contrast. Then the thickness of the identified flake was measured using non-contact AFM in a Park XE-100. Since it is well documented that a 1-2nm interfacial layer of residue can exist between the substrate and 2D material[32], the AFM results were corroborated by looking at the Raman signature of the flake. Monolayer to 5-layer flakes of $MoTe_2$ can be distinguished by the ratio of the Raman $A_{1g}$ and $B_{2g}$ peaks [23]. The combined AFM + Raman was important for unambiguous determination of thin flakes. However, for thicker flakes, the AFM result was used and the 1-2nm uncertainty was wrapped up in the error.

Storage of Ionic Liquid

DEME-TFSI was purchased from Iolitec in bulk and stored in a nitrogen glove box. Before transferring into the glovebox, the ionic liquid was first dried in a vacuum oven at 80C for 24 hours to remove any water that might have been introduced during the shipping process. Small volumes of DEME-BF$_4$ were removed from the glovebox on an experiment by experiment basis.

Preparation of the Gating Cell

The gating cell was built to be easily implemented, disposable, and cheap. The cell consists of two electrodes, a 3D-printed exterior well, and a 3D-printed interior well capped at one end with a quartz coverslip. At the end of the exfoliation process mentioned above, the MoTe$_2$ flake rests on polished, p++ doped silicon which acts both as a substrate and the electrode. The 3D-printed exterior well is epoxied (Lysol 1H) onto the p++ silicon such that the MoTe$_2$ flake of interest is in the center of the well. The epoxy cures overnight at room temperature in a nitrogen box to minimize oxygen exposure to the MoTe$_2$. The counter-electrode or gate is a gold wire that is coiled to rest inside of the exterior well and propped up such that it does not touch the p++ Si substrate. Right before the in-situ Raman experiment, ~300 μL of DEME-TFSI is added to the well to fill it half way. The interior well is then placed inside of the exterior well, displacing a large volume of the ionic liquid and allowing a liquid-free path to the MoTe$_2$ flake for the Raman laser. The quartz coverslip of the interior well presses up against the MoTe$_2$ flake and Si substrate, forming a thin interfacial layer of ionic liquid that minimizes unwanted scattering but can still rearrange its ions to react to the applied electric field. After measurement, the gold wire and interior well are washed with repeated rinses of isopropyl alcohol and water and can be used again.

The height of the gating cell was chosen to be smaller than the working distance of the 100X LWD objective used in the Raman experiments. The dimensions of the gold wire were chosen to have a higher surface area then the area of the p++ Silicon electrode.

Oxygen Exposure of MoTe$_2$

MoTe$_2$ is known to be moderately oxygen sensitive on the order of hours to days[33]. The MoTe$_2$ exfoliation tapes and prepares substrates are stored in a nitrogen box to minimize exposure. The estimated total air exposure during experimental preparation is 3-4 hours. No signs of oxidation peak have been found in Raman measurements prior the gating.

In-situ Raman measurements

The in-situ Raman gating experiments were performed using a Horiba Labram Evolution with a 633nm laser and a 100X LWD (0.6 NA) objective. All experiments were done in ambient conditions. A laser power of 0.52 mW was chosen to minimize thermal heating (see Supplementary Figure 8). A Keithley 2400 was used at the voltage source and contacts were made by means of XYZ micromanipulator probes and alligator clips. An 1800 g/mm was used to maximum peak resolution and the Raman region of 50 to 350 wavenumbers was studied. Measurements times and accumulations varied but were generally a 10 second acquisition with 6 accumulations or a 30 second acquisition with 3 accumulations. When the voltage was changed, the ions were given 30 to 60 s to equilibrate before the Raman measurement was started.

Polarized Raman Measurements

Polarized-resolved Raman was achieved by using a 633nm half-wave plate mounted before the objective to rotate the incident linearly polarized light. No additional polarizers were used after the objective and the Raman signal was collected in reflected mode by the Silicon CCD.

DFT Calculations

All DFT calculations were performed using the Vienna Ab Initio Simulation Package[34–37], version 5.4.4. All calculations use PAW pseudopotentials to treat the core electrons[38,39]. The lithium atom on bilayer $MoTe_2$ calculations were performed using 4 formula units of $MoTe_2$ (an orthorhombic cell consisting of 2 formula units on each layer). To accurately treat the potential between layers which has a van der Waals component, we use the SCAN + rVV10[40,41] functional. For these calculations, an energy cutoff of 400 eV is used. Atom positions are allowed to relax in the bilayer. We use tetrahedral integration of the Brillouin zone with a smearing parameter of 0.05 eV and an energy cutoff of $10^{-4}$ eV.

For calculations of vibrational modes, we use the PBE GGA functional[42], along with the Phonopy [43] package to analyze the vibrational modes. A 16 f.u. cell of $MoTe_2$ with a 400 eV cutoff and $10^{-5}$ eV energy convergence threshold are used, along with Gaussian smearing with a 0.05 eV smearing width. Phonon modes are computed using the built-in density functional perturbation theory (DFPT) method implemented in VASP.

All input files for calculations can be found at: https://github.com/rehnd/MoTe2-MultiLayerPhaseChange

**ASSOCIATED CONTENT**

**Supporting Information**
The Supporting Information in available free of charge on XX at DOI: XXXXX. The SI includes nomenclature of phonon modes, comparison of 2H and 1T' Raman peaks in literature, discussion of anomalous Raman intensity enhancement, peak centers and linewidths of Raman data, estimations of laser heating, investigation of intercalation using XRD and photoluminescence, and phase change in 3-4L flake.

**Acknowledgements**

Part of this work was performed at the Stanford Nano Shared Facilities (SNSF), supported by the National Science Foundation under award ECCS-1542152. This work was also partially supported by NSF Grants No. EECS-1436626 and No. DMR-1455050, Army Research Office Grant No. W911NF-15-1-0570, Office of Naval Research Grant No. N00014-15-1-2697, and a seed grant from Stanford System X Alliance. This work was supported in part by the US Army Research Laboratory, through the Army High Performance Computing Research Center, Cooperative Agreement No. W911NF-07-0027. The authors would like to thank A.R. Chew, Y. Li, and O.B. Aslan for helpful discussion. The authors acknowledge support from B.A. Reeves with in-situ photoluminescence.  D.Z. acknowledges support by the National Science Foundation Graduate Research Fellowship under Grant No. DGE-1656518.

# Supplementary Information:

# Reversible Electrochemical Phase Change in Monolayer to Bulk MoTe$_2$ by Ionic Liquid Gating

**Dante Zakhidov[1], Daniel A. Rehn[2,3], Evan J. Reed[1], Alberto Salleo[1,*]**

[1]Department of Materials Science and Engineering, Stanford University, Stanford, CA 94305, USA
[2]Department of Mechanical Engineering, Stanford University, Stanford, CA 94305, USA
[3]Los Alamos National Laboratory, Los Alamos, NM 87545, USA

*Email: asalleo@stanford.edu

**Table of Contents:**

SI Table 1: Note on nomenclature of phonon modes

The assignment of Raman phonon modes is based on group theory as dictated by crystal symmetry. Bulk 2H MoTe$_2$ is part of the P6$_3$/mmc space group with a D$_{6h}$ point group. However, due to loss of translational symmetry in the out-of-plane direction in the few-layer limit, there is symmetry breaking depending on the parity of the number of layers. Flakes with an odd number of layers have a D$_{3h}$ point group and flakes with an even number of layers have a D$_{3d}$ point group.[1] This changes the nomenclature of the irreducible representations used to define the point group. A summary of the nomenclature in the various parity conditions is shown below.[2] Similarly, for 1T' MoTe$_2$ in the few-layer limit, an odd number of flakes has a C$_{2h}$ point group while flakes with even number of layers have a C$_s$ point group.[3] The point group is written in Schoenflies notation and space group is written in international short symbol or Hermann-Mauguin notation.

Since in our phase change work, we cannot identify the number of layers transformed to the 1T', we assign all 1T' Raman modes using the bulk notation convention for consistency. In the case of a monolayer 2H to 1T' transition, we use the monolayer notation. For the 2H modes, we label the modes appropriately based on thickness determined from a combination of AFM, Raman, and optical transmission.

| Point Group (Space Group) | 2H | 1T' |
|---|---|---|
| Odd # Layers | D$_{3h}$ (P6m2) | C$_{2h}$ (P2$_1$/m) |
| Even # Layers | D$_{3d}$ (P3m1) | C$_s$ (Pm) |
| Bulk | D$_{6h}$ (P6$_3$/mmc) | C$_{2h}$ (P2$_1$/m) |

| Number of Layers | 2H ~170cm$^{-1}$ | 2H ~234 cm$^{-1}$ | 2H ~290 cm$^{-1}$ | 1T' ~77cm$^{-1}$ | 1T' ~94 cm$^{-1}$ | 1T' ~163 cm$^{-1}$ |
|---|---|---|---|---|---|---|
| Odd | A$_1$' | E' | A$_2$'' | A$_g$ | B$_g$ | A$_g$ |
| Even | A$_{1g}$ | E$_g$ | A$_{1g}$ | A' | A'' | A' |
| Bulk | A$_{1g}$ | E$_{2g}$ | B$_{2g}$ | A$_{1g}$ | B$_{1g}$ | A$_{1g}$ |

SI Table 2: Comparison of 2H and 1T' Raman Peak Modes Reported in Literature

Table 2a shows 1T' $MoTe_2$ peak positions as reported in literature. Values reported for "HQ Graphene 1T'" were measured in-house. Table 2b shows 1T' $MoTe_2$ that was phase changed from a starting 2H flake. This work is the first to report the presence of 1T' $A_g$ and $B_g$ peaks. Bolded values represent peaks that have qualitatively high intensities in the Raman spectra.

| Table 2A: 1T' Raman peak positions from literature | $A_g$ (A') | $B_g$ (A'') | $B_g$ (A'') | $A_g$ (A') | $A_g$ (A') | $A_g$ (A') | $B_g$ (A'') | $A_g$ (A') | $A_g$ (A') |
|---|---|---|---|---|---|---|---|---|---|
| Beams, et al. (calculated)[3] | 78 | 94 | 107 | 111 | 128 | 163 | 192 | 249 | 260 |
| HQ Graphene 1T' (experimental, bulk) | **77.5** | 94.7 | 107.2 | | 128.5 | **163.8** | | | 260 |
| Empante et al. (CVD, 1L)[4] | **80** | 85 | 102 | 112 | 126 | **162** | | | |
| Naylor et al. (CVD, 1L)[5] | | | | 112 | 127 | **161** | | 252 | 269 |
| Park et al. (CVD, ~20nm)[6] | | | 107 | | 126.9 | **163** | | | 256.1 |
| Zhou et al. (CVD, ?)[7] | **78** | 93 | | 110 | 128 | **162** | 189 | | 258 |

| Table 2B: 1T' Raman peak positions after 2H -> 1T' phase change | $A_g$ (A') | $B_g$ (A'') | $B_g$ (A'') | $A_g$ (A') | $A_g$ (A') | $A_g$ (A') | $B_g$ (A'') | $A_g$ (A') | $A_g$ (A') |
|---|---|---|---|---|---|---|---|---|---|
| Song et al. (strain phase change, >20nm)[8] | | | | 120 | 140 | | | | |
| Cho et al. (laser phase change, bulk)[9] | | | | 124 | **138** | | | | |
| Tan et al. (laser phase change, bulk) | | | | **125** | 140 | | | | |
| Wang et al. (electrostatic, 1L)[10] | | | | | | **167.5** | | | |
| **This work (electrostatic, electrochemical, 1L to bulk)** | 78 | 90 | | | | **168** | | | |

Peak Centers and Linewidths for the Bilayer and Bulk-Like MoTe$_2$ Gating Experiments

In SI Figures 1 and 2, we plot and tabulate the peak center and peak full-width half-maximum (FWHM) of the 2H A$_{1g}$, 2H E$_{2g}$, and 1T' A$_{1g}$ for the bulk-like and bilayer experiments respectively. This quantitative data corresponds to the Raman spectra in Figure 2 of the paper. The peaks were fitted by Lorentzian functions and the error bars in the graphs come from the error associated with the fitting function.

For the 60nm case, we see that the 2H E$_{2g}$ peak is best fit by two Lorentzian functions at and around the transition voltage. For example, at 4.2V the E$_{2g}$ peak is split with a peak center at 234.3 and 231.9 cm$^{-1}$ and linewidths of 2.97 and 14.96 cm$^{-1}$ respectively. The red-shifted, broad linewidth peak is interpreted as a defective 2H signal which most likely comes from un-changed 2H on the surface of the flake that borders growing 1T' domains. The pristine 2H E$_{2g}$ component comes from underlying layers. At higher voltages, the pristine signal disappears which could be interpreted as multiple layers becoming defective and the excitation laser no longer being able to reach underlying pristine layers due to the absorption of the material. We find that that 2H signals return to similar peak centers and linewidths but not the exact same, which indicates there have been irreversible changes to the MoTe$_2$.

In the bilayer case, we also get defective 2H E$_{2g}$ but in this instance, the process is more gradual. There is no split peak, most likely because the charge affects both layers and there are no underlying pristine 2H layers. In the bilayer case, the 2H peak centers and linewidths nearly return to their starting values.

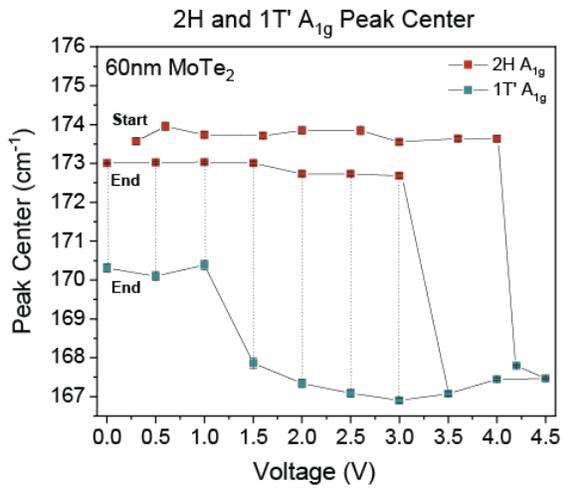
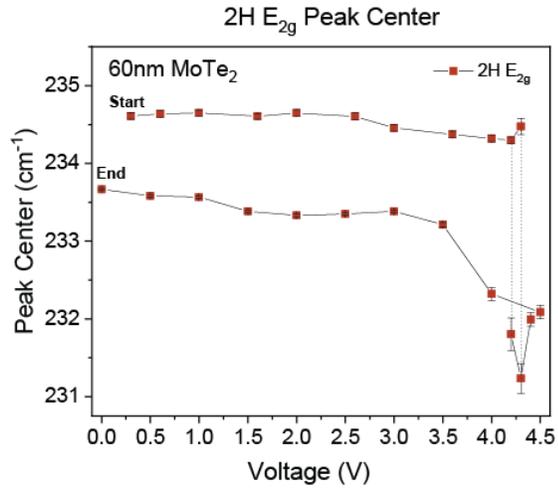
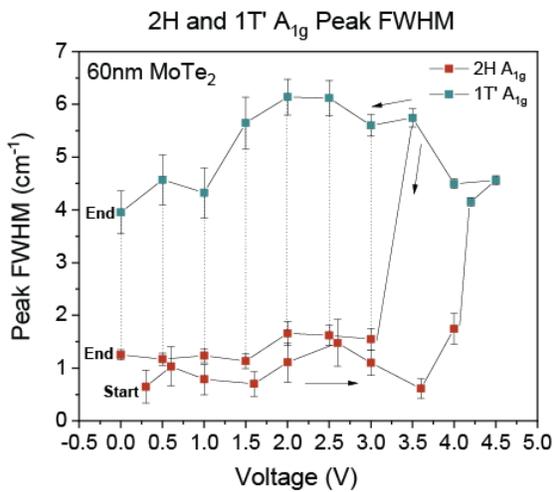
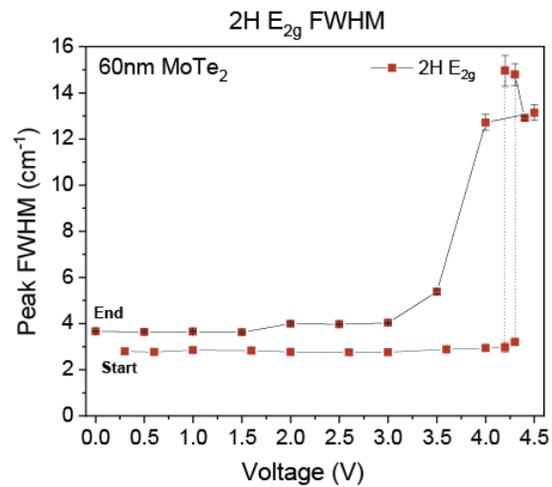

| 60 nm | $A_{1g}$ Peak Center (cm$^{-1}$) | | $A_{1g}$ Peak FWHM (cm$^{-1}$) | |
|---|---|---|---|---|
| Voltage (V) | 2H | 1T' | 2H | 1T' |
| 0.3 | 173.57 | | 0.65 | |
| 0.6 | 173.95 | | 1.03 | |
| 1 | 173.74 | | 0.79 | |
| 1.6 | 173.71 | | 0.70 | |
| 2 | 173.85 | | 1.11 | |
| 2.6 | 173.85 | | 1.48 | |
| 3 | 173.55 | | 1.10 | |
| 3.6 | 173.64 | | 0.61 | |
| 4 | 173.64 | | 1.74 | |
| 4.2 | | 167.79 | | 4.15 |
| 4.5 | | 167.46 | | 4.56 |
| 4 | | 167.44 | | 4.50 |
| 3.5 | | 167.07 | | 5.74 |
| 3 | 172.68 | 166.90 | 1.55 | 5.60 |
| 2.5 | 172.73 | 167.08 | 1.62 | 6.12 |
| 2 | 172.73 | 167.34 | 1.66 | 6.14 |
| 1.5 | 173.00 | 167.86 | 1.13 | 5.65 |
| 1 | 173.03 | 170.39 | 1.24 | 4.33 |
| 0.5 | 173.03 | 170.10 | 1.17 | 4.57 |
| 0 | 173.00 | 170.31 | 1.25 | 3.95 |

| 60 nm | $E_{2g}$ Peak Center (cm$^{-1}$) | | $E_{2g}$ Peak FWHM (cm$^{-1}$) | |
|---|---|---|---|---|
| Voltage (V) | 2H | 2H Split | 2H | 2H Split |
| 0.3 | 234.61 | | 2.79 | |
| 0.6 | 234.64 | | 2.77 | |
| 1 | 234.65 | | 2.84 | |
| 1.6 | 234.61 | | 2.82 | |
| 2 | 234.65 | | 2.77 | |
| 2.6 | 234.61 | | 2.76 | |
| 3 | 234.46 | | 2.76 | |
| 3.6 | 234.37 | | 2.88 | |
| 4 | 234.32 | | 2.94 | |
| 4.2 | 234.30 | 231.80 | 2.97 | 14.96 |
| 4.3 | 234.48 | 231.23 | 3.20 | 14.79 |
| 4.5 | | 232.09 | | 13.14 |
| 4 | | 232.32 | | 12.71 |
| 3.5 | 233.21 | | 5.38 | |
| 3 | 233.38 | | 4.03 | |
| 2.5 | 233.35 | | 3.97 | |
| 2 | 233.33 | | 3.99 | |
| 1.5 | 233.38 | | 3.62 | |
| 1 | 233.57 | | 3.65 | |
| 0.5 | 233.58 | | 3.64 | |
| 0 | 233.67 | | 3.66 | |

**SI Figure 1. Raman Peak Values for Bulk-Like 60nm MoTe$_2$ Flake.** The peak centers and linewidths are both plotted and tabulated for a full gating cycle from 0V -> 4.5V -> 0V. The dotted lines connecting two points represent a co-existence of the 2H and 1T' phase or two different 2H peaks. Arrows are added to clarify the gating direction. The tables contain the same information as the graphs but with the values listed for ease of reference.

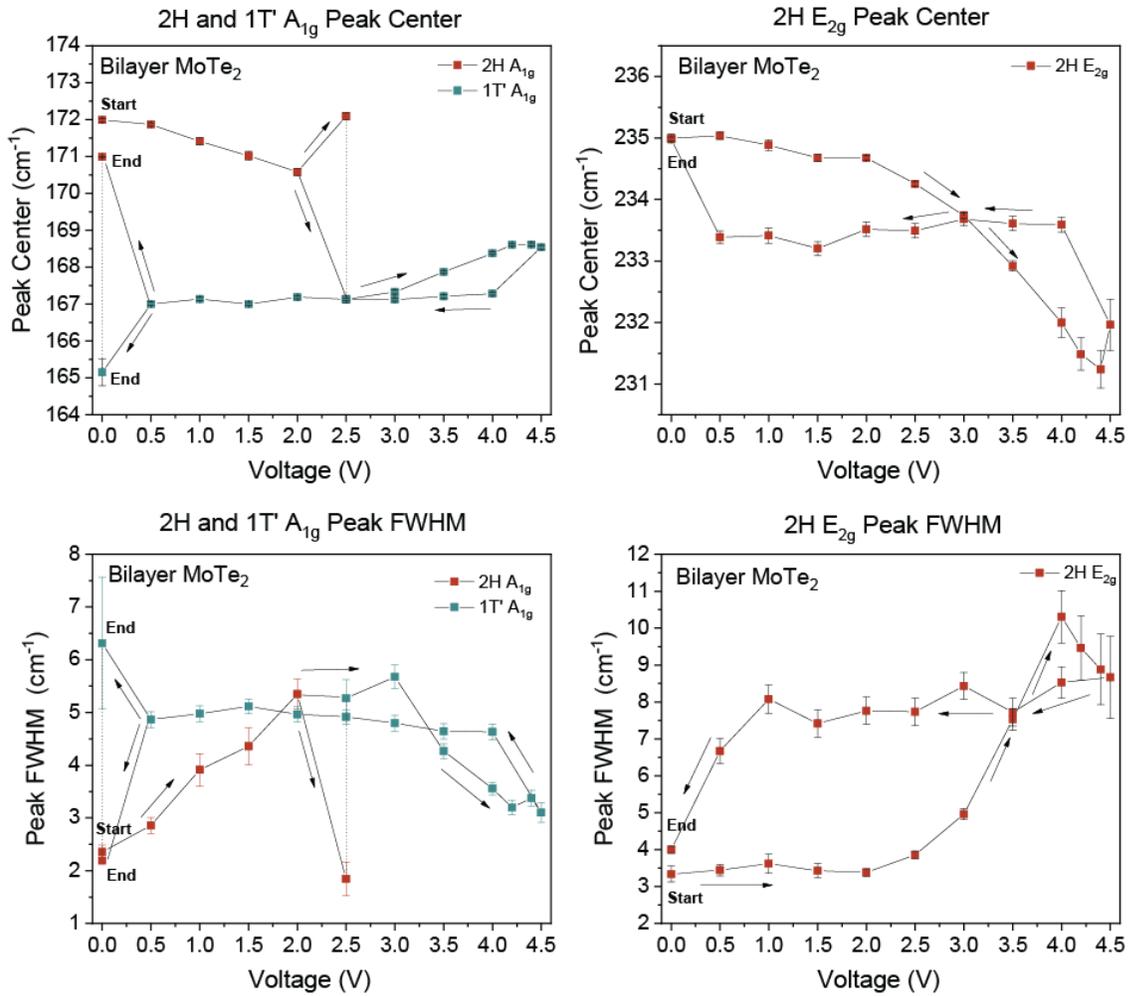

| Bilayer Voltage (V) | $A_{1g}$ Peak Center (cm$^{-1}$) 2H | 1T' | $A_{1g}$ Peak FWHM (cm$^{-1}$) 2H | 1T' |
|---|---|---|---|---|
| 0 | 171.99 | | 2.35 | |
| 0.5 | 171.87 | | 2.86 | |
| 1 | 171.41 | | 3.91 | |
| 1.5 | 171.02 | | 4.36 | |
| 2 | 170.58 | | 5.35 | |
| 2.5 | 172.09 | 167.13 | 1.84 | 5.27 |
| 3 | | 167.33 | | 5.68 |
| 3.5 | | 167.87 | | 4.27 |
| 4 | | 168.38 | | 3.56 |
| 4.2 | | 168.61 | | 3.20 |
| 4.4 | | 168.61 | | 3.38 |
| 4.5 | | 168.54 | | 3.10 |
| 4 | | 167.28 | | 4.63 |
| 3.5 | | 167.21 | | 4.64 |
| 3 | | 167.12 | | 4.80 |
| 2.5 | | 167.13 | | 4.92 |
| 2 | | 167.19 | | 4.97 |
| 1.5 | | 167.00 | | 5.12 |
| 1 | | 167.13 | | 4.98 |
| 0.5 | | 167.00 | | 4.87 |
| 0 | 170.99 | 165.15 | 2.19 | 6.31 |

| Bilayer Voltage (V) | $E_{2g}$ Peak Center (cm$^{-1}$) 2H | 2H Split | $E_{2g}$ Peak FWHM (cm$^{-1}$) 2H | 2H Split |
|---|---|---|---|---|
| 0 | 234.99 | | 3.34 | |
| 0.5 | 235.04 | | 3.44 | |
| 1 | 234.88 | | 3.62 | |
| 1.5 | 234.68 | | 3.43 | |
| 2 | 234.68 | | 3.38 | |
| 2.5 | 234.26 | | 3.86 | |
| 3 | 233.74 | | 4.96 | |
| 3.5 | 232.92 | | 7.53 | |
| 4 | 232.00 | | 10.31 | |
| 4.2 | 231.49 | | 9.46 | |
| 4.4 | 231.24 | | 8.88 | |
| 4.5 | 231.96 | | 8.67 | |
| 4 | 233.59 | | 8.53 | |
| 3.5 | 233.61 | | 7.72 | |
| 3 | 233.68 | | 8.44 | |
| 2.5 | 233.49 | | 7.73 | |
| 2 | 233.52 | | 7.76 | |
| 1.5 | 233.20 | | 7.42 | |
| 1 | 233.41 | | 8.08 | |
| 0.5 | 233.39 | | 6.67 | |
| 0 | 234.98 | | 4.00 | |

**SI Figure 2. Raman Peak Values for Bilayer MoTe$_2$ Flake.** The peak centers and linewidths are both plotted and tabulated for a full gating cycle from 0V -> 4.5V -> 0V. The dotted lines connecting two points represent a co-existence of the 2H and 1T' phase or two different 2H peaks. Arrows are added to clarify the gating direction. The tables contain the same information as the graphs but with the values listed for ease of reference.

Anomalous Raman Enhancement Effect

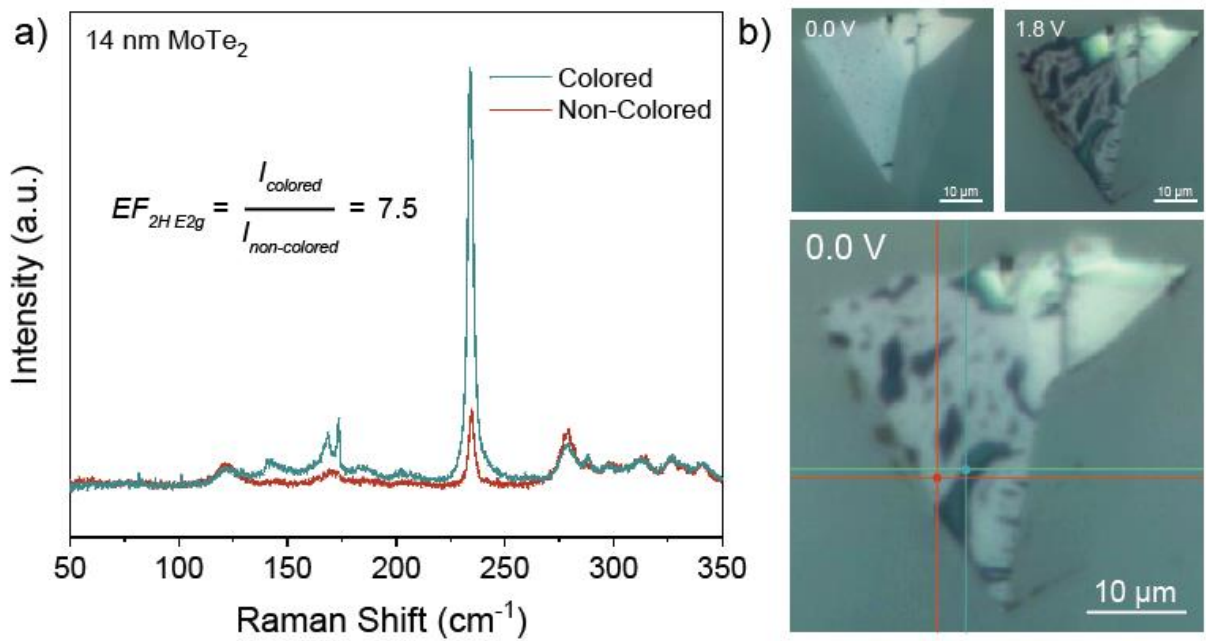

**SI Figure 3. Visualization of Raman Enhancement. a)** Comparing Raman signals taken on a colored area vs a non-colored area on a MoTe$_2$ flake after cycling back to 0.0 V. The enhancement factor, EF, is taken by fitting the 2H E$_{2g}$ peaks and dividing the intensities. **b)** Optical images chronologically showing the flake throughout the gating process. At 1.8V, we see an overall darkening of the flake in addition to large patches of colored areas. The final image at 0.0 V has color-coded crosshairs depicting the exact areas of the measurement.

We repeatedly but inconsistently observe a Raman enhancement effect that occurs at or after the transition voltage in the gating process. In this experiment, a bulk flake was cycled from 0 V to 1.8 V and then back to 0 V. At 1.8 V we observe 2H to 1T' phase change by Raman spectroscopy in addition to striated color changes on the flake. Upon removing the voltage, the overall darkening of the flake goes away but some large colored areas remain. Measuring the Raman spectrum on a colored-area and a non-colored area (SI Figure 3a), we find that the colored-area has retained the 1T' phase and has an intensity enhancement factor of around 7.5. The non-colored area does not show the 1T' signature. SI Figure 3b shows the optical images through the gating cycle with the final 0.0V highlighting the exact measurement positions, color-coded to match the spectrum in SI Figure 3a.

The Raman enhancement is associated with color changes on the surface of the flake but we find that it is not due to the 1T' phase itself. We can see this from Figure 2c in the paper, where we get the 2H to 1T' transition in bilayer MoTe$_2$ but don't see any enhancement of the Raman signal. To further test the possibility of a Raman enhancement from a 1T' top layer, we create a 1T' on 2H heterostructure using a PDMS stamp transfer method with an XYZ micromanipulator stage. The assembly of the heterostructure was motivated by a recent report showing a SERS effect where 1T' MoTe$_2$ was used as a substrate[11]. This heterostructure is different with 1T' on top of a 2H substrate but accurately matches our experimental conditions. The 1T' region is estimated to be 2-3L based on optical transmission contrast and the 2H region is 5L as we can see by the Davydov splitting in the Raman spectrum. Looking at the heterostructure, we see there is no intensity increase in the Raman spectrum from having 1T' on top of 2H. In fact, the intensity of both materials is attenuated.

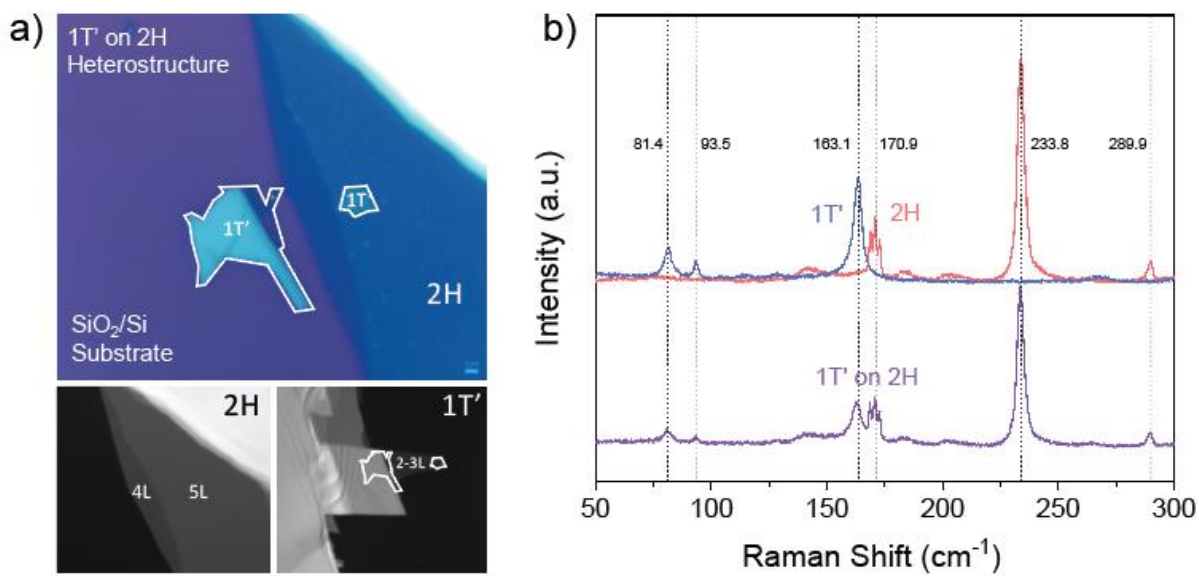

**SI Figure 4. 1T' on 2H Heterostructure. a)** Optical images showing the creation of the 1T' on 2H heterostructure. The colored image shows the final product. The black and white images show the 2H and 1T' MoTe$_2$ flakes that were selected for the heterostructure. During the transfer process, only a portion of the 1T' stayed on the 2H which is outlined in white. **b)** Raman spectrum of the 2H, 1T', and 1T' on 2H

Raman intensity enhancement is generally attributed to surface-sensitive interactions and phenomena is called surface-enhanced Raman scattering (SERS). There are two general mechanisms which explain the enhancement:

(1) Electromagnetic Mechanism: Enhancement from the excitation of localized surface plasmons from a physisorbed metal (Au,Ag) on top of the material of interest. Gives large enhancement factors of $10^5$ to $10^9$.
(2) Chemical Mechanism: Enhancement from formation of charge-transfer complexes due to a species that has formed a chemical bond with the surface of the material. Gives smaller enhancement factors of $10^1$ to $10^3$.

We speculate we are observing a chemical SERS effect due to the relatively low enhancement factor. Furthermore, the formation of Te metal in our conditions suggest strong chemical interactions between the ionic liquid and the MoTe$_2$ surface. Further work is needed to clarify the nature of this Raman enhancement effect and find its origin, which is beyond the scope of this manuscript.

Investigation of Intercalation as a Potential Explanation for Multi-Layer Phase Change

Intercalation of 2D layered materials has been widely studied in the case of liquid-phase exfoliation of bulk 2D materials[12], creation of intercalated superlattices[13], and most relevantly, $MoS_2$ phase change by means of lithium intercalation[14]. As a result, it was important to test whether the $DEME^+$ cation of the DEME-TFSI ionic liquid was intercalating in our $MoTe_2$ and whether it was contributing to the phase change process.

Intuitively, lithium atoms are small while the DEME cation is much larger, which would make intercalation seem unfeasible. However, in the work on intercalated superlattices, Cetyltrimethylammonium Bromide (CTAB) was used as the electrolyte and found to stably intercalate a variety of 2D materials. The XRD spectrum after CTAB intercalation showed a 115% increase in the interlayer distance. The $CTA^+$ cation is similar in size to the $DEME^+$ cation which makes the possibility of intercalation worth testing.

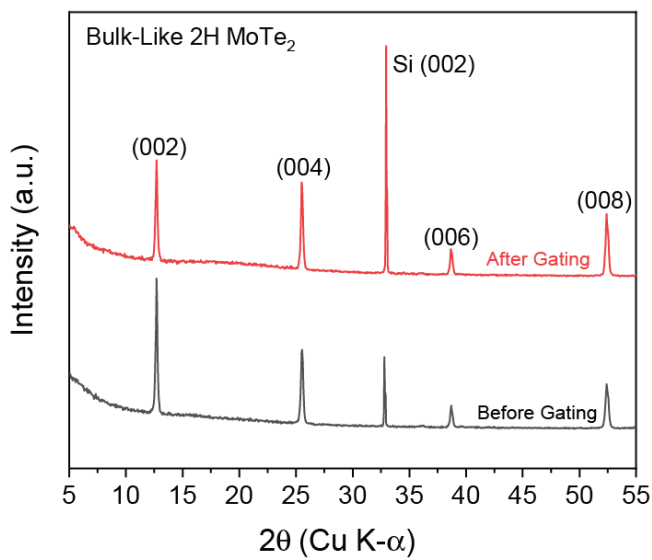

**SI Figure 5. XRD of 2H MoTe₂ Before and After Gating.** Specular scan of a high-density exfoliation of 2H $MoTe_2$ flakes on a p++ Si substrate. All flakes were simultaneously gated using the gating structure in methods. The gating cell and ionic liquid were removed and then measured in XRD ex-situ.

SI Figure 5 shows the XRD spectra of a large density of bulk-like 2H $MoTe_2$ flakes exfoliated on a p++ substrate before and after gating. We see no shift in the out-of-plane peaks of the $MoTe_2$ which would indicate no intercalation. However, the measurements were done ex-situ and we cannot rule out the possibility that the cations de-intercalated before the XRD measurement was taken.

## Video of Simultaneous Phase Change in 4nm and 12nm regions of a MoTe$_2$ Flake

Supplementary Video 1 shows two MoTe$_2$ flakes through an optical video feed used during our in-situ Raman gating experiments. The flake in the center of the video has 4nm, 12nm, and 37nm thick regions as described in Figure 4 of the paper. At this point in the experiment, the flake has already been phase change cycled between 2.4V and 0V several times. At the start of the video, there was 0V applied to the flakes. At 3 seconds into the video, 2.4V is applied to the gating cell and we see coloration across the flake which coincides with the 2H to 1T' phase change. The video highlights the simultaneous phase change in two different thicknesses and lack of phase change in bulkier regions. It also shows the kinetics and striated nature of the coloration. Note that in the audio of the video, it was mentioned that -2.4V is applied to the flake but this was due to the configuration of electrodes at the time. Based on the gating cell schematic shown in Figure 1 of the paper, we applied 2.4V which injected electrons into the MoTe$_2$.

## Phase Change in 3L or 4L MoTe$_2$ and Evidence Against the 1T' Peak being a Davydov Peak

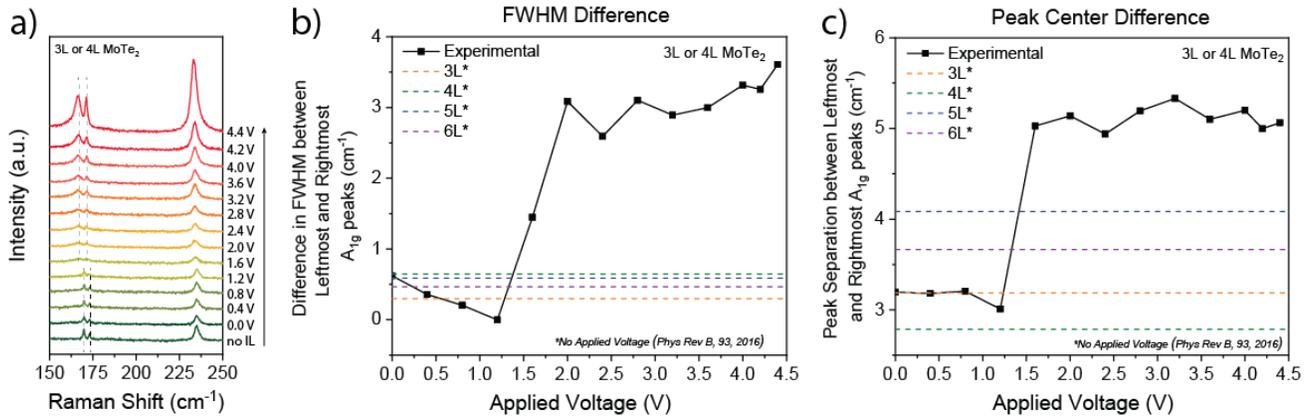

**SI Figure 6. Phase Change in Few Layer MoTe$_2$ with Comparisons Against Davydov Splitting. a)** Raman gating cycle for a 3L or 4L flake with the 2H to 1T' phase transition at 1.6 V. The 2H A$_{1g}$ doublet at no applied voltage is characteristic of 3 or 4L MoTe$_2$. **b)** Comparing the FWHM difference between the left and right-most peaks around ~170 cm$^{-1}$ to distinguish between Davydov splitting and the 2H to 1T' phase transition. **c)** Comparing peak position difference.

In the few-layer case of group VI-TMDS, the out-of-plane degenerative phonon modes split due to interlayer interactions in a phenomenon known as Davydov splitting[16]. In 2H MoTe$_2$, the A$_{1g}$ peak exhibits Davydov splitting between 3-10L. An example of Davydov splitting can be seen in SI Figure 6a where two 2H A$_{1g}$ peaks are visible in pristine MoTe$_2$ in a non-gating condition.

Upon gating the flake, we see the phase change at 1.6V, with the creation of the new 1T' peak. Given the proximity of the phase changed 1T' peak to the original 2H peak, it is important to distinguish the 1T' peak from a Davydov peak.

We extract the linewidth and peak center of Davydov split peaks from literature[16] and compare them to our experimental data in SI Figures 6b and 6c. The linewidths of Davydov-split peaks are nearly identical with little variation. We see that the linewidth of the phase-changed 1T' peak is significantly larger than the 2H peak which clearly marks it apart from a Davydov split peak. Likewise, when looking at peak centers, we see that the peak center difference between the Davydov-split peaks is smaller than the peak center difference between the co-existing 2H and 1T' peaks. This comparison shows that the new peak from the gating process is indeed a 1T' peak and not a Davydov-split peak.

Projection Analysis of 1T' $A_{1g}$ Mode under the Presence of Vacancies

To explore the effect of Te vacancy formation on the Raman spectra of 1T' MoTe2, we perform DFT-based calculations of the vibrational modes of pristine and ~3% Te-deficient MoTe2 (see Methods section of the main paper for computational details). We look specifically at the $A_g$ mode of monolayer MoTe$_2$, which in the pristine case has a computed frequency of 162.58 cm-1 (SI Figure 7b). In the case of a single Te vacancy in a 16 f.u. cell (SI Figure 7a), the mode displays a very small frequency shift to 162.84 cm-1 (SI Figure 7c). In this case, the mode displacement vectors largely retain the same values as in the pristine case (compare panels b and c). In order to investigate character of the vibrational modes in the Te-deficient case, we perform a projection of every mode onto the pristine 162.58 cm-1 mode. The value computed is

$$\text{proj} = \frac{u_\nu^v \cdot u_{163}^0}{||u_\nu^v||_2 ||u_{163}^0||_2} \qquad (1)$$

Here $u_\nu^v$ is the mode corresponding to frequency $\nu$ in the vacancy (superscript $v$) case, while $u_{163}^0$ is the 162.58 cm$^{-1}$ mode in the zero vacancy (superscript 0) case. The projection data shown in panel d shows that only the 162.84 cm-1 mode shown in panel c bears a great deal of similarity to the 162.58 cm-1 mode in panel b (the modes match by roughly 90% according to this metric). Therefore, we conclude that the vibrational modes will not change significantly upon Te vacancy formation.

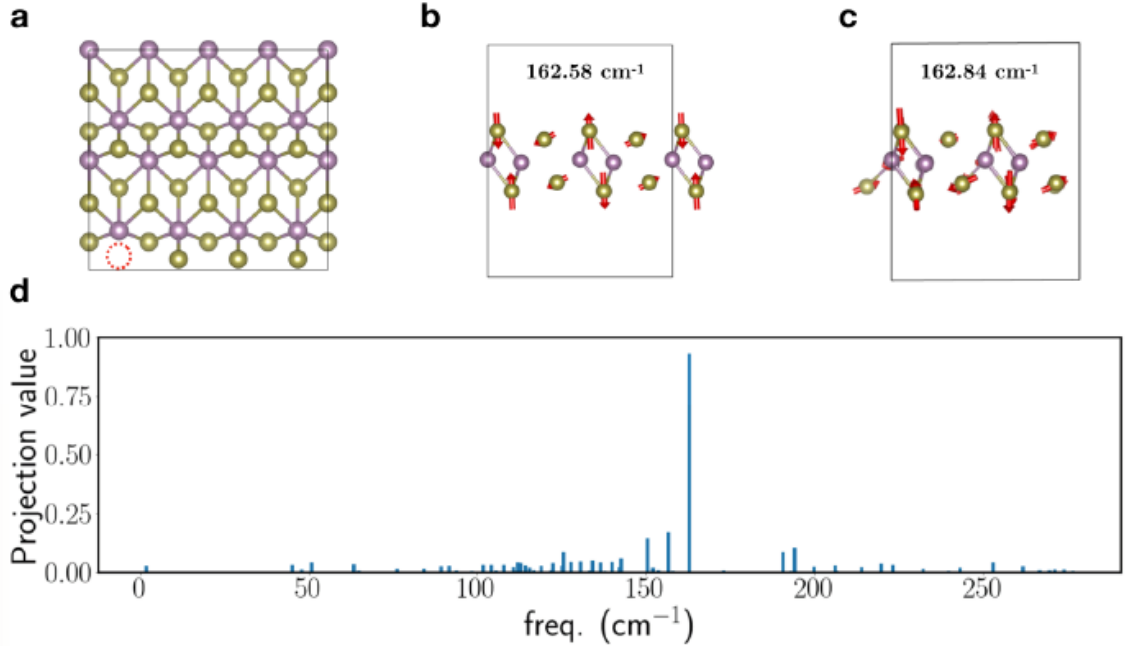

**SI Figure 7. (a)** the original 16 f.u. cell of 1T'-MoTe2 with a single Te vacancy created in the red dashed circle. **(b)** shows the Raman active mode at 162.58 cm$^{-1}$ in the pristine case. **(c)** shows the most similar mode to the 162.58 cm$^{-1}$ mode in the pristine case when the vacancy shown in (a) is present. This mode has frequency 162.84 cm$^{-1}$. **(d)** shows the projection values of all modes in the structure of (a) onto the mode shown in (b), as computed by Eq. 1, clearly showing that the mode shown in (c) is highly similar to the mode shown in (b), while all other modes are dissimilar.

Minimizing Effect of Raman Laser Heating on Phase Change Processes

The laser power of the excitation laser needs to be carefully selected to minimize heating contributions to the gate-induced phase change. In VI-TMD literature, laser powers for a 633nm laser of ~0.25 mW to 1 mW are used in Raman experiments because those powers were found to not alter the Raman spectrum upon extended and repeated exposure[8,10]. We similarly use a laser power of 0.52 mW to probe our 2H and 1T' phases for all experiments.

In SI Figure 8, we estimate the effect of heating due to various laser powers based on the stokes/anti-stokes ratio and 2H $A_{1g}$ and $E_{2g}$ peak shift. Since the stokes and anti-stokes Raman signals are based on phonon populations in the ground and excited states, the ratio of the signals is related to the temperature of the material through the equation in SI Figure 8b[17]. The stokes/anti-stokes ratio used in these approximations is based on the peak height and not the integrated peak-intensity. We choose the peak height as an indicator of peak intensity because at higher laser powers, the peaks were found to be asymmetrical with sizeable shoulders, making accurate peak fitting difficult. The temperature can also be estimated based on the peak shift and is tabulated in SI Figure 8c. We use temperature coefficients of -0.0055 and -0.0175 cm$^{-1}$/K for the 2H $A_{1g}$ and $E_{2g}$ peaks respectively[18].

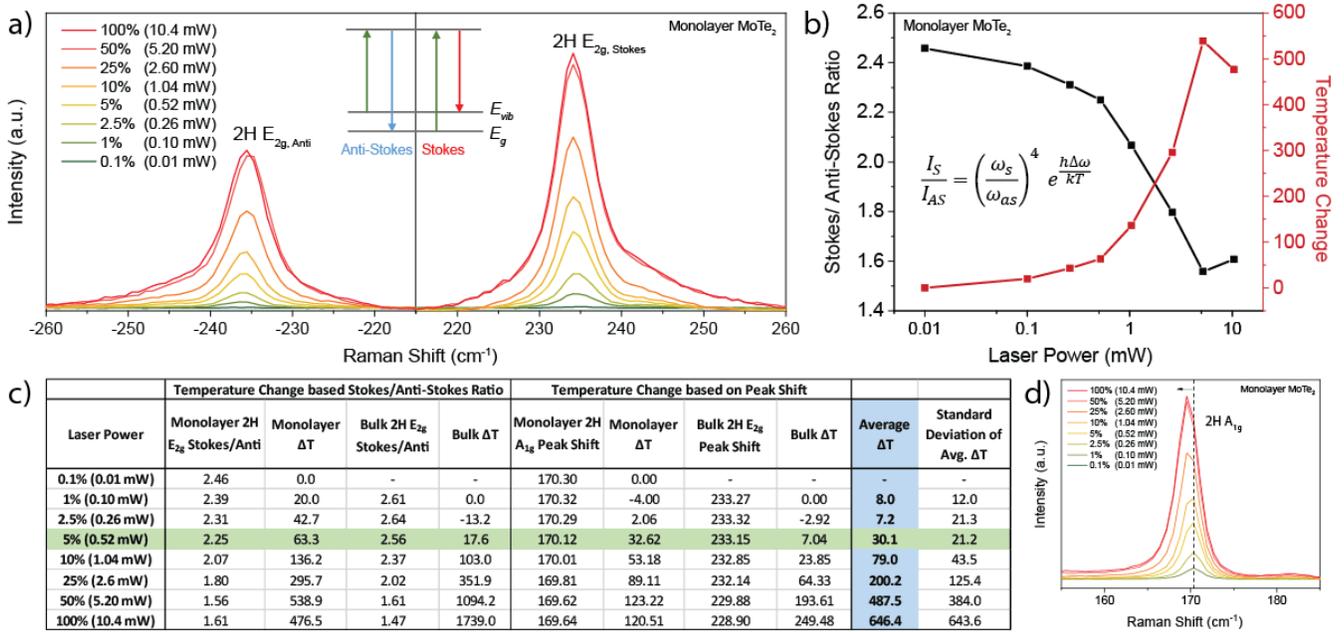

**SI Figure 8. Estimations of Heating from Raman Laser. a)** Stokes and Anti-Stokes 2H $E_{2g}$ peaks for a bulk-like MoTe$_2$ flake of varying laser powers of a 633nm laser. **b)** Using the stokes/anti-stokes ratio to estimate heating from the data in part a. **c)** Tabulating various methods of estimating temperature from Raman peaks. The data from part a,b is listed first followed by an identical analysis for a bulk-like flake. The temperature is then estimated based on peak shift values of monolayer and bulk-like MoTe$_2$. The average change in temperature and standard deviation of the methods is also shown. We use a laser power setting of 0.52 mW to minimize the impact of heating in our experiments. **d)** Peak shift in the monolayer 2H $A_{1g}$ peak upon heating with varying laser powers.

Based on those methods, we find the average change in temperature when using a 0.52 mW 633 nm laser is 30 degrees with a standard deviation of 21 degrees. Upon using higher laser powers, we see that the temperature change can be quite significant but also note that the standard deviation is on the order of the change. We find that these techniques become inaccurate at large temperatures and we speculate it's due to other phenomena that can be altering the Raman peak position or intensity. For example, vacancy defect creation or the formation of Te metal. In the end, the 5% laser power (0.52 mW) was used in all experiments to minimize heating effects from the laser. No changes to the Raman spectrum induced by laser heating were observed when using this laser power.